\documentclass[pra, aps, 10pt, twocolumn, floatfix,superscriptaddress]{revtex4-1} 

\usepackage {graphicx}
\usepackage {amssymb}
\usepackage {amsmath}
\usepackage {verbatim}
\usepackage{color}
\usepackage[normalem]{ulem}

\begin{document}

\title{Generation of an ultrahigh-repetition-rate optical half-cycle pulse train in the nested quantum wells} 

\author{Mikhail Arkhipov}

\affiliation{St. Petersburg State University, Universitetskaya nab. 7/9, St. Petersburg 199034, Russia}

\affiliation{Ioffe Institute, Politekhnicheskaya str. 26, St. Petersburg 194021, Russia}

\author{Anton Pakhomov}

\affiliation{St. Petersburg State University, Universitetskaya nab. 7/9, St. Petersburg 199034, Russia}

\author{Rostislav Arkhipov}

\affiliation{St. Petersburg State University, Universitetskaya nab. 7/9, St. Petersburg 199034, Russia}

\affiliation{Ioffe Institute, Politekhnicheskaya str. 26, St. Petersburg 194021, Russia}

\author{Nikolay Rosanov}

\affiliation{St. Petersburg State University, Universitetskaya nab. 7/9, St. Petersburg 199034, Russia}

\affiliation{Ioffe Institute, Politekhnicheskaya str. 26, St. Petersburg 194021, Russia}

\begin{abstract}
We propose a simple quantum system, namely, a nested quantum-well structure, which is able to generate a train of half-cycle pulses of a few-fs duration, when driven by a static electric field. We theoretically investigate the emission of such a structure and its dependence on the parameters of the quantum wells. It is shown that the production of a regular output pulse train with tunable properties and the pulse repetition frequencies of tens of THz is possible in certain parameter ranges. We expect the suggested structure can be used as an ultra-compact source of subcycle pulses in the optical range.
\end{abstract}

\maketitle

The generation of ultra-short pulses down to few- and subcycle ones, especially in the optical range, has become one of the major research areas in the modern optics over last decades inspired by a variety of promising applications in the control of different ultrafast processes in matter \cite{Krausz, Keller, Mourou_Nobel, Biegert, Midorikawa, Mondal, Xue_2022, Hui}. Specifically, half-cycle pulses 
were shown to provide efficient excitation of different quantum systems on the timescales well below the periods of the respective resonant transitions \cite{Dimitrovski_PRL, Dimitrovski_PRA,Arkhipov2019_OL, Arkhipov_JL_2021, Tumakov_PRA, Pakhomov_2022}. 
Still the shorter pulses are needed, the more difficult it is to find a convenient source of such ultra-short pulses.

The available commercial lasers with passive mode-locking has made it possible to routinely generate  trains of few-cycle pulses of sub-10-fs duration, such as the well-known Ti:sapphire laser \cite{Rafailov}. 
However, for the production of single-cycle and subcycle pulses in the optical range other methods are required.

Trains of even shorter pulses down to sub-fs duration can be  produced using the high-harmonic generation \cite{Goulielmakis_HHG} or the interaction of strong two-colour infrared laser pulses with atoms \cite{persson2006generation}.  
A few methods have been proposed so far for the generation of isolated half-cycle unipolar and quasi-unipolar pulses, e.g. the excitation of a foil target by intense femtosecond pulses \cite{Wu, Xu, Eliasson, Eliasson_2, Eliasson_3}, the transformation of a bipolar pulse in a non-equilibrium photo-ionized plasma \cite{bogatskaya2021new, Bogatskaya, Bogatskaya_2}, 
the optical attosecond pulse synthesis \cite{Hassan}, through cascaded processes in plasma \cite{Shou_Mourou} or the formation of unipolar half-cycle solitons in different nonlinear media \cite{kaplan1995electromagnetic, kaplan1997electromagnetic, Kalosha, Parkhomenko, Vysotina, Song, Kozlov, Sazonov2018, Leblond2019, Sazonov_UP_1, Sazonov_UP_2}. 
Still all these methods require exceedingly bulky and complicated experimental setups. 
Thus, finding more simple, compact and convenient sources of subcycle pulses and pulse trains in the optical range remains the challenging and attracting issue for researchers.

\begin{figure}[tpb]
\centering
\includegraphics[width=1.\linewidth]{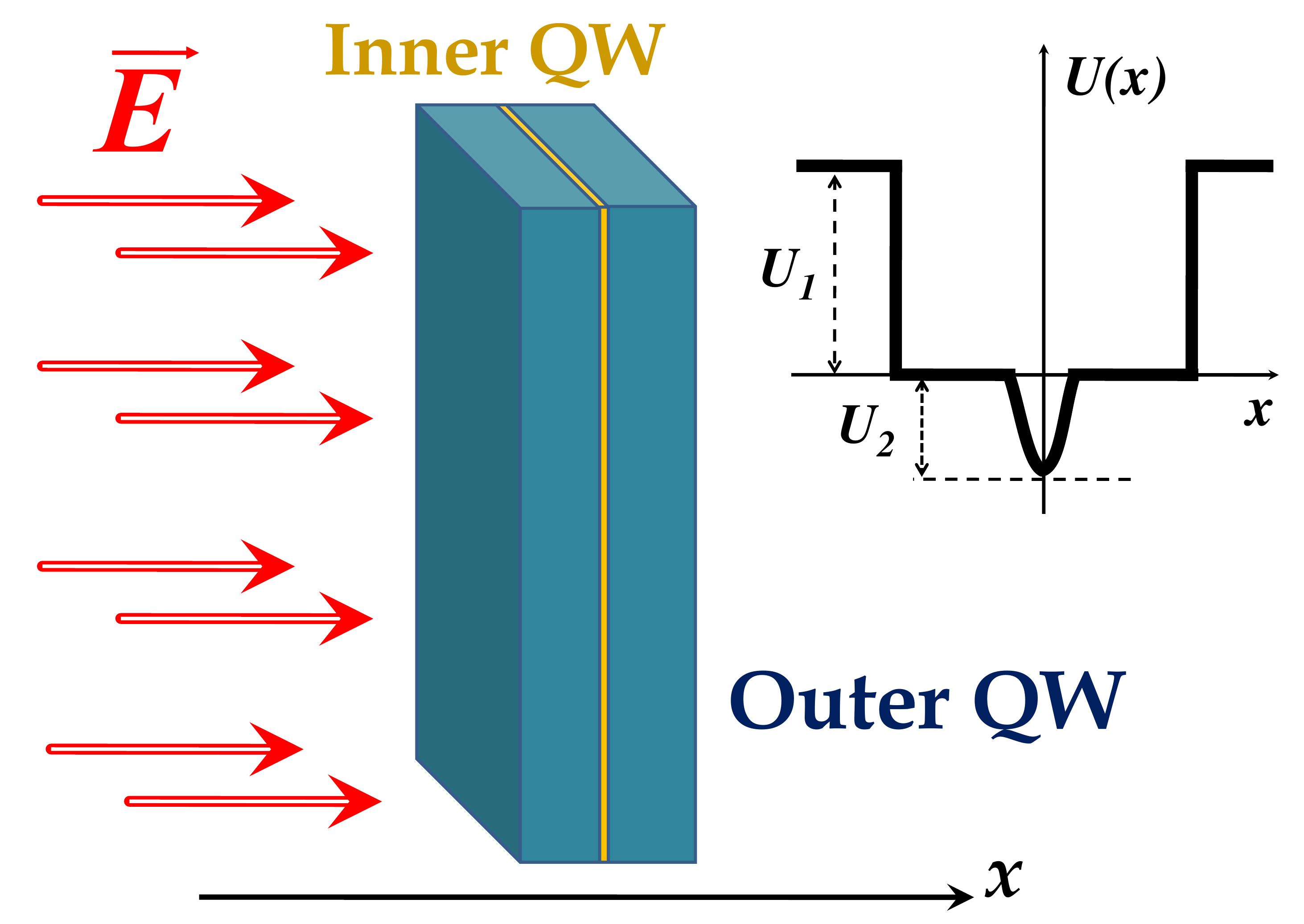}
\caption{(Color online) The considered nested quantum-wells structure driven by an external homogeneous static electric field $\vec E$; the inset shows the potential energy profile of the outer quantum well of the depth $U_1$ with an embedded inner quantum well of the depth $U_2$.}
\label{fig1}
\end{figure}

In this Letter we propose a simple quantum system, which can generate a train of half-cycle pulses of several fs in duration, when driven by an applied static homogeneous electric field. Our system represents a pair of nested quantum wells (QW), as schematically shown in Fig.~\ref{fig1}.

In the case of a single quantum well the wave function of an electron inside it would occupy in the ground state the whole width of the well. Applying an external field weak enough to prevent the ionization could cause some trembling of the electron inside the well, which would be however barely detectable. Our idea is to introduce an extra inner quantum well inside the outer quantum well. As the result, with the proper choice of the inner QW's depth the electron in the ground state gets trapped by the inner QW, what allows to localize the ground state wave function. Now applying an external field strong enough to cause the electron to leave the inner QW but weak enough to keep it inside the outer quantum well can lead to the electron oscillations inside the outer well. Such oscillating electron when getting reflected at the outer quantum well's edge at each round-trip can be expected to emit an ultra-short subcycle pulse, so that a regular pulse train should be  eventually obtained.

In order to model the dynamics of the considered system, we use the standard time-dependent Schr{\"o}dinger equation for the electron wave function $\psi(x,t)$ \cite{Landau}:
\begin{equation}
i\hbar \frac{\partial \psi}{\partial t} = \Big[ \hat H_0 + \hat V(x,t) \Big] \psi,
\label{SE}
\end{equation}
where the intrinsic Hamiltonian $\hat H_0$ of the considered one-dimensional system  is given as:
\begin{equation}
\nonumber
\hat H_0 \psi = - \frac{\hbar^2}{2 m_e} \frac{\partial^2 \psi}{\partial x^2}  + U(x) \psi
\label{H0}
\end{equation}
with the effective electron mass $m_e$ and the intrinsic potential energy profile of the quantum wells $U(x)$. The interaction potential with the external electric field is given as:
\begin{equation}
\nonumber
V(x,t) = e E(t) x
\label{Vx}
\end{equation}
with the electron charge $e$. The applied electric field was taken static, i.e.:
\begin{equation}
\nonumber
E(t) = E \cdot \Theta(t),
\label{Et}
\end{equation}
with the constant amplitude $E$ and the Heaviside step function $\Theta$.

\begin{figure}[tpb]
\centering
\includegraphics[width=1.\linewidth]{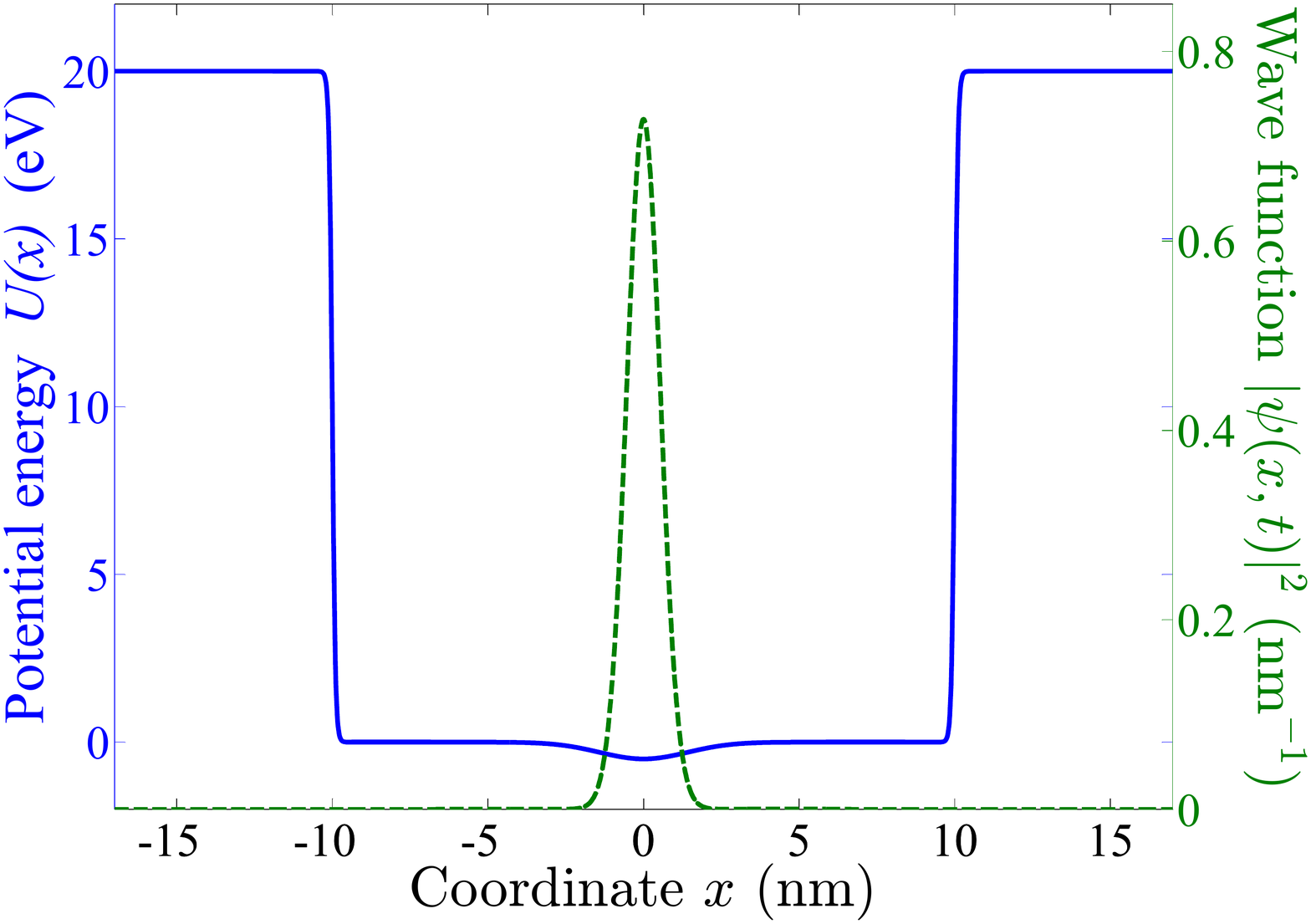}
\caption{(Color online) The potential energy profile of the quantum wells given by Eqs.~\eqref{Ux}-\eqref{Udefect} and the wave function of the respective ground state; the parameters are: $U_1$ = 20 eV, $U_2$ = 0.5 eV, $L_1$ = 10 nm, $L_2$ = 2 nm, $D_1$ = 0.05 nm.}
\label{fig2}
\end{figure}

The intrinsic potential energy consists of 2 parts, namely the outer and inner quantum wells:
\begin{equation}
U(x) = U_{\text{outer}} (x) + U_{\text{inner}} (x).
\label{Ux}
\end{equation}
We take here a finite outer quantum well of close-to-rectangular shape with sharp edges, specifically of the following Fermi-Dirac-type profile:
\begin{equation}
U_{\text{outer}} (x) = U_1 \Big( 1 - \frac{1}{1 + e^{\frac{|x| - L_1}{D_1}}} \Big)
\label{U_QW}
\end{equation}
with the half-width of the outer QW $L_1$ and the thickness of the edges $D_1$. For the inner QW we use the Gaussian shape:
\begin{equation}
U_{\text{inner}} (x) = -U_2 \ e^{-x^2/L_2^2}
\label{Udefect}
\end{equation}
where $L_2$ is the inner QW's width, and we assume that the inner QW just slightly disturbs the potential energy profile, namely:
\begin{equation}
U_2 \ll U_1.
\label{shallow}
\end{equation}
An example of the potential energy profile is plotted in Fig.~\ref{fig2} together with the respective wave function of the ground state. One can see that even though the inner QW's depth is almost two orders of magnitude smaller than the outer well's depth, it still allows to efficiently localize the ground state within the inner QW instead of occupying the whole width of the outer QW.

\begin{figure}[tpb]
\centering
\includegraphics[width=1.\linewidth]{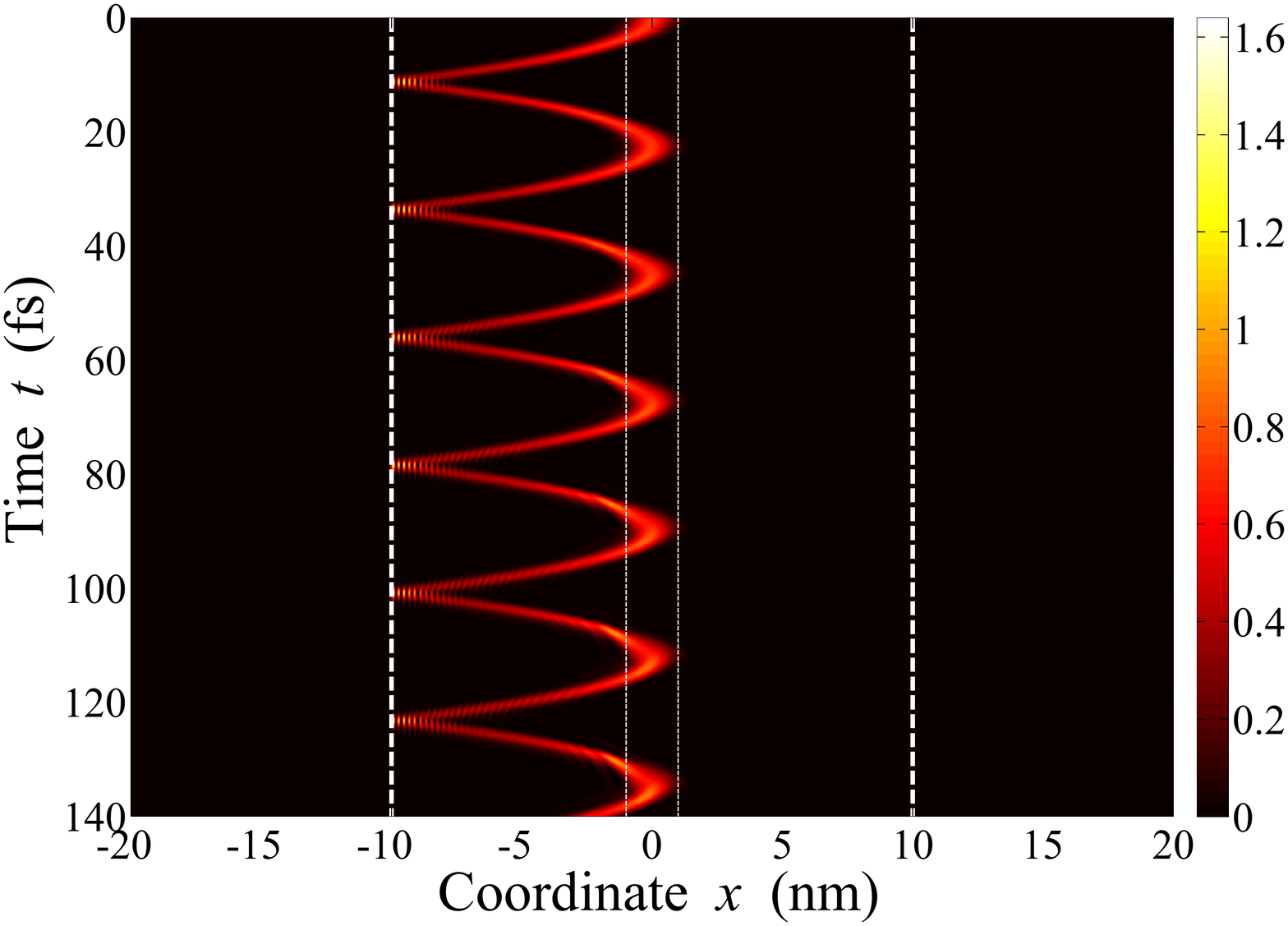}
\caption{(Color online) The spatio-temporal evolution of the wave function $|\psi(x,t)|^2$ (expressed in nm$^{-1}$) for the considered quantum-well structure; the parameters are: $U_1$ = 20 eV, $U_2$ = 0.5 eV, $E$ = 10$^9$ V/m, $L_1$ = 10 nm, $L_2$ = 2 nm, $D_1$ = 0.05 nm. The thick dashed white lines indicate the boundaries of the outer quantum well, while the thin dashed white lines outline the boundaries of the inner quantum well (at the $e^{-1}$ level). Initially the system was in the ground state.}
\label{fig3}
\end{figure}

The system emission is provided by the second-order temporal derivative of the induced medium polarization:
\begin{equation}
E_{\text{emit}} (t) \sim \frac{\partial^2 P}{\partial t^2}  \Big|_{t - R/c},
\label{emission}
\end{equation}
where $R$ is the distance between the emitter and the detector. The induced polarization for the considered system can be obtained as:
\begin{equation}
P(t) = -e N \int_{-\infty}^{+\infty} \psi(x,t) \ x \  \psi^*(x,t) dx,
\label{polariz}
\end{equation}
with the electron spatial density $N$. The second-order temporal derivative of the induced medium polarization Eq.~\eqref{polariz} is therefore proportional to the electron acceleration $\ddot x (t)$, which is given as \cite{Landau}:
\begin{eqnarray}
\nonumber
\ddot x (t) &=& \int_{-\infty}^{+\infty} \psi(x,t) \Big( - \frac{\nabla U(x)}{m_e} - \frac{\nabla V(x,t)}{m_e} \Big) \psi^*(x,t) dx  \\
\nonumber
&=& \frac{1}{m_e} \int_{-\infty}^{+\infty} \psi(x,t) \Big( -\nabla U(x) - e E(t) \Big) \psi^*(x,t) dx. \\
\label{acceleration}
\end{eqnarray}

\begin{figure}[tpb]
\centering
\includegraphics[width=1.\linewidth]{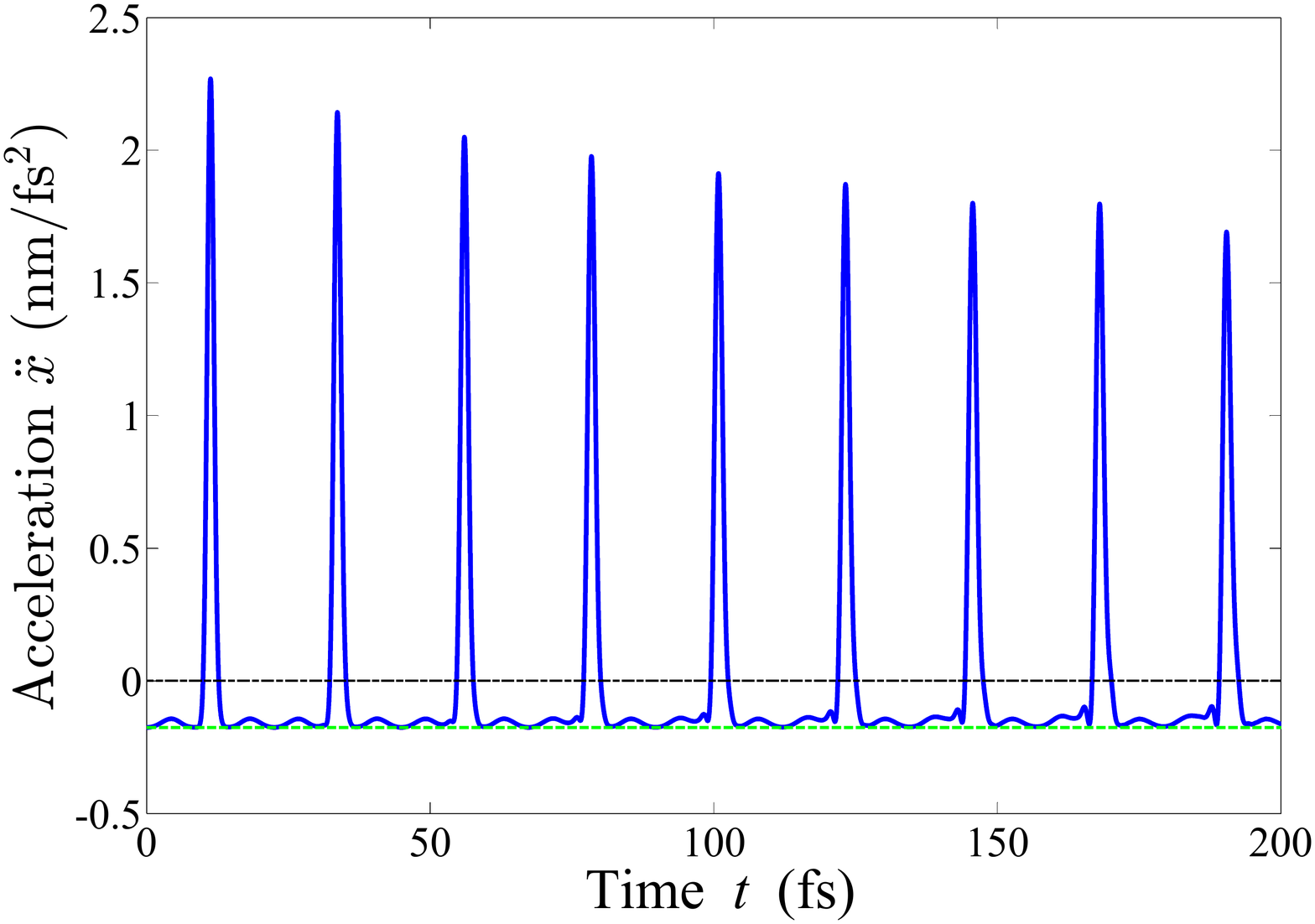}
\caption{(Color online) The electron acceleration in the quantum wells vs. time for the parameters of Fig.~\ref{fig3}; the dashed green line corresponds to the value $-e E / m_e$, i.e. the acceleration of a free electron in an external electric field.}
\label{fig4}
\end{figure}

The Schr{\"o}dinger equation Eq.~\eqref{SE} was solved numerically using the pseudo-spectral Split-step Fourier method (SSFM) \cite{Agrawal}.
An example of the arising dynamics is shown in Figs.~\ref{fig3}-\ref{fig4} for the parameters from Fig.~\ref{fig2}. Fig.~\ref{fig3} demonstrates the spatial and temporal evolution of the squared wave function of the system $|\psi(x,t)|^2$. As can be seen, the electron rapidly leaves its initial localized state in the inner QW and undergoes the periodic oscillations inside the outer QW. At the left boundary of the outer QW the electron wavepacket is each time reflected and returns back to the inner QW. The resulting trajectory is close to exact parabolas typical for the particle motion under an external field in the classical mechanics.

The corresponding time trace of the electron acceleration Eq.~\eqref{acceleration} is plotted in Fig.~\ref{fig4}. One can see that upon each reflection from the boundary the electron is to emit a pronounced half-cycle burst according to Eq.~\eqref{emission}. The pulse duration at half-maximum is just around 1.5 fs. In between the successive reflections the electron acceleration is almost constant and equal to $-e E/ m_e$, with only slight deviations caused by the presence of the inner QW. The resulting emission in Fig.~\ref{fig4} therefore appears to be a sequence of half-cycle bursts against the background of a much weaker field of the opposite polarity.

Let us find out the range of the electric field strength for the applied external field, where the pulse train can be generated. To assure that the electron accelerated by the applied external field can not get over the energy barrier and leave the outer quantum well it is required that:
\begin{equation}
E (L_1 + S_{\text{inner}}) \lesssim U_1,
\label{threshold1}
\end{equation}
where $S_{\text{inner}}$ is the shift of the inner QW with respect to the center of the outer quantum well in the direction of the external electric field. Eq.~\eqref{threshold1} implies that the energy transferred to the electron by the external field does not exceed the depth of the outer QW. At the same time the electric field has to be strong enough to free the electron trapped in the ground state onto the inner QW. The binding force of the inner QW can be estimated as $ - \nabla U_{\text{inner}} (x) \sim U_2 / L_2$. Therefore the applied field can kick the electron out of the inner QW as long as the driving field strength exceeds this binding force. Together with the condition Eq.~\eqref{threshold1} we thus end up with the following range of the field strength for the efficient generation of a half-cycle pulse train:
\begin{equation}
\frac{U_2}{L_2}  \lesssim  E  \lesssim \frac{U_1}{L_1 + S_{\text{inner}}}.
\label{E_range}
\end{equation}
As the performed numerical studies exhibit, the derived range Eq.~\eqref{E_range} indeed turns out to be in good agreement with the modeling results.

The underlying mechanism of the half-cycle pulses generation can be described as follows. First, the external electric field kicks the electron out of the inner QW. Provided that the condition Eq.~\eqref{E_range} is obeyed, one gets the whole range of the electric field strength, so that the applied field promotes the electron to leave the inner QW but is not strong enough to drive the electron out of the outer QW. As the result, the electron starts to oscillate inside the outer QW. During each round-trip of its motion the electron gets accelerated by the external static field towards one of the boundaries of the outer QW, then bounces off this boundary and changes the direction of its velocity, after that slows down by the external static field until the electron reaches its initial position at the inner QW and so on. During the motion away from the boundaries of the outer QW the electron acceleration is almost constant and proportional to the applied field. At the same time nearby the boundary the electron gets rapidly reflected, so that its acceleration becomes much larger in amplitude and has the opposite sign. The emitted field according to Eq.~\eqref{emission} simply follows the time dependence of the electron acceleration.

The peak pulse amplitude does not stay constant, but rather slowly decays with time
due to the electron scattering both at the inner QW and at the edge of the outer QW upon the reflections from it. Despite this, in our numerical simulations the electron oscillations and the resulting emission of well-pronounced subcycle pulses, e.g. for the parameters of Figs.~\ref{fig3}-\ref{fig4}, last over at least a few ps.

The obtained amplitude of the generated half-cycle pulses gets larger as the electric field larger accelerates the electron wavepacket until its collision with the boundary of the outer QW. Therefore, in order to increase the output pulse amplitude one has to increase the product of the electric field strength $E$ and the distance $L_1 + S_{\text{inner}}$. However, as the condition Eq.~\eqref{threshold1} states, in this case the outer QW's depth $U_1$ has to be also accordingly increased for the described phenomena to be preserved.

The arising motion the electron undergoes can be well described by the laws of the classical mechanics, namely as the motion of a mass point $m_e$ driven by a constant force $-e E$. The resulting pulse repetition period can be therefore calculated as:
\begin{equation}
T_{r} = 2 \sqrt{\frac{2 (L_1 + S_{\text{inner}}) m_e}{e E}}.
\label{T_r}
\end{equation}
For instance, inserting the parameters from Figs.~\ref{fig2}-\ref{fig4} with $S_{\text{inner}} = 0$ into Eq.~\eqref{T_r} one finds for the pulse repetition frequency $f_r = (T_{r})^{-1} = 46.9$ THz. So high repetition rates deep into the THz range by orders of magnitude exceed typical values for laser sources. Our system thus enables generation of half-cycle pulse trains with ultra-high repetition rates.

\begin{figure}[tpb]
\centering
\includegraphics[width=1.\linewidth]{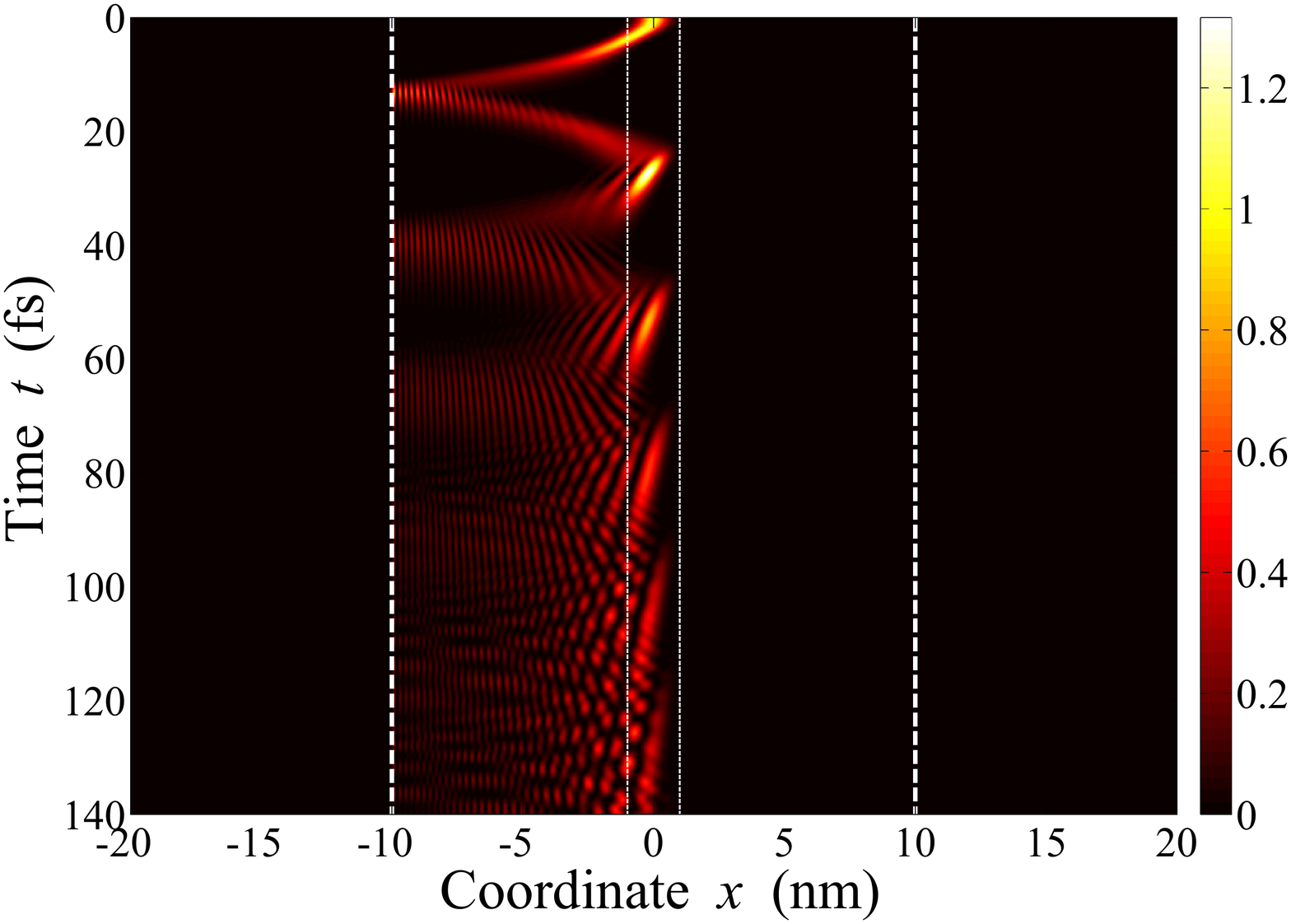}
\caption{(Color online) The spatio-temporal evolution of the wave function $|\psi(x,t)|^2$ (expressed in nm$^{-1}$) for the considered quantum-well structure; the parameters are: $U_1$ = 20 eV, $U_2$ = 2 eV, $E$ = 10$^9$ V/m, $L_1$ = 10 nm, $L_2$ = 2 nm, $S_{\text{inner}} =0$, $D_1$ = 0.05 nm. The thick dashed white lines indicate the boundaries of the outer quantum well, while the thin dashed white lines outline the boundaries of the inner quantum well (at the $e^{-1}$ level). Initially the system was in the ground state.}
\label{fig5}
\end{figure}

Let us address now the conditions on the structure parameters, which enable obtaining of a half-cycle pulse train. As the conditions on the applied electric field are provided by Eqs.~\eqref{threshold1}-\eqref{E_range}, we are mainly interested in the suitable values of the variables $U_1$, $U_2$, $L_1$ and $L_2$. The main criteria for the existence of the described dynamical regime is however still provided by the same expression Eq.~\eqref{E_range} meaning that the values of $U_1$, $U_2$, $L_1$ and $L_2$ have to be chosen in such a way that the left-hand side of Eq.~\eqref{E_range} is significantly smaller than the right-hand side.

\begin{figure}[tpb]
\centering
\includegraphics[width=1.\linewidth]{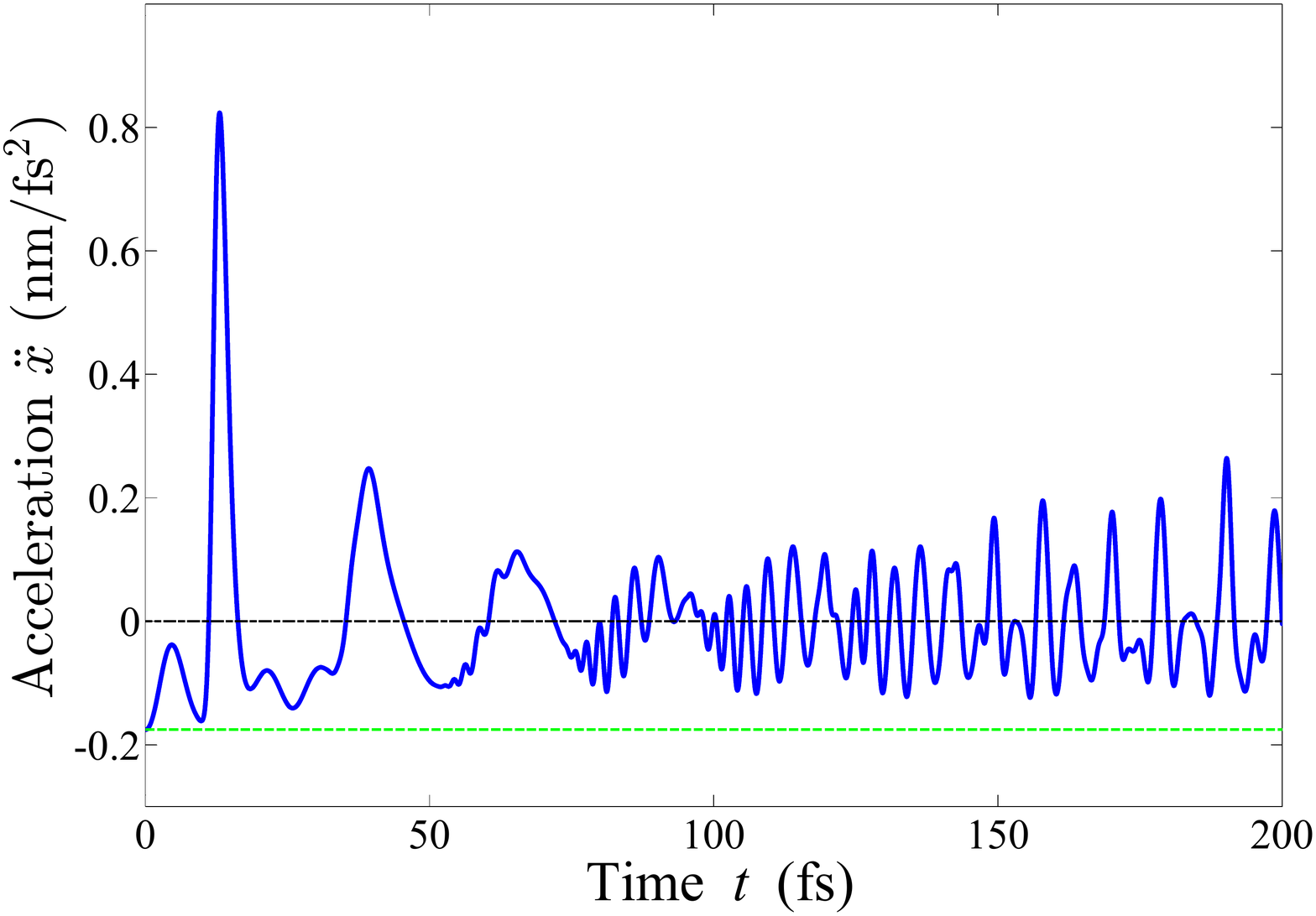}
\caption{(Color online) The electron acceleration in the quantum wells vs. time for the parameters of Fig.~\ref{fig5}; the dashed green line corresponds to the value $-e E / m_e$, i.e. the acceleration of a free electron in an external electric field.}
\label{fig6}
\end{figure}

Besides that, as our numerical simulations yield, even provided that Eq.~\eqref{E_range} is obeyed, the ratio between the values $U_1$ and $U_2$ largely affects the quality of the produced pulse train. As the inner QW should only provide a minor deviation of the potential energy profile according to  Eq.~\eqref{shallow}, this condition has to be made more precise. Actually if the inner QW's depth $U_2$ is too small as compared to $U_1$, the inner QW fails to localize the electron ground state within it, so that the electron wave function occupies almost the whole width of the outer QW $L_1$ and applying the electric field does not result in the pronounced oscillations inside the outer QW.

Inversely, if the inner QW's depth $U_2$ becomes large enough as compared to $U_1$, the electron experiences significant scattering on the inner QW upon its oscillations in the outer quantum well. Such scattering leads to the rapid broadening of the wave function and very irregular and disordered output emission instead of regular pronounced bursts like in Fig.~\ref{fig4}. An example of such dynamics is illustrated in Figs.~\ref{fig5}-\ref{fig6}, where we used the same parameters as in Figs.~\ref{fig2}-\ref{fig4} except for the value of $U_2$ increased to $U_2 = 2$ eV. One can easily see both the rapid scattering of the electron wave function on the inner QW in Fig.~\ref{fig5} and the respective random output emission in Fig.~\ref{fig6}. Hence, only in a certain range of the values of the ratio $U_2 / U_1$ a proper high-quality output pulse train is to be detected. Specifically, for the fixed value of $U_1 = 20$ eV, a well-pronounced regular half-cycle pulse train was found to arise for the $U_2$ values in the range $U_2 \approx 0.1 - 1$ eV.

In conclusion, we have theoretically demonstrated the half-cycle pulses generation in a simple quantum system, representing just a pair of nested quantum wells. In the absence of an external field, the electron in such a structure gets trapped by the inner QW and its wave function becomes localized within this inner QW. When an external static electric field is applied, the electron can leave the inner QW and starts moving inside the outer quantum well. If the depth of the outer quantum well is large enough, as given by the condition Eq.~\eqref{threshold1}, the electron can not escape the outer quantum well and periodically oscillates inside, while producing a half-cycle burst upon each reflection from the edge of the outer quantum well. Such a structure allows generation of half-cycle pulse trains with the pulse repetition rates of the order of tens of THz. The proposed system with nested QWs could be fabricated using semiconductor multiple-quantum-well heterostructures \cite{Kulbachinskii_2012}.  
We believe that our findings are able to provide a noticeable contribution to the creation of feasible ultra-compact sources of subcycle pulses in the optical range.

\section*{Acknowledgments}

Russian Science Foundation, project 21-72-10028.
The authors are grateful to Dr. Ihar Babushkin for fruitful discussions on the paper findings over last years.

\bibliography{UP_library}

\end{document}